\documentclass[a4paper]{ISOStyle}
\usepackage{epsfig, graphics}

\begin{document}

\title{Searching for AGN signatures in Mid-IR Spectra: The case of NGC1068}

\vspace{15mm}

\author{O. Laurent\inst{1,2}, I.F. Mirabel\inst{2,3}, V. Charmandaris\inst{4}, 
E.~Le~Floc'h\inst{2,5}, D. Lutz\inst{1} \and R. Genzel\inst{1}} 

\institute{
  Max-Planck-Institut f\"ur extraterrestrische Physik, Postfach 1603,
  85740 Garching, Germany 
  \and CEA/DSM/DAPNIA Service d'Astrophysique F-91191 Gif-sur-Yvette,
  France
  \and Instituto de Astronom\'\i a y F\'\i sica del Espacio. cc 67,
  suc 28. 1428 Buenos Aires, Argentina
  \and  Astronomy Department, Cornell University, IRS Science Center, 
  Ithaca NY, 14853, USA
  \and European Southern Observatory, Karl-Schwarzschild-Str, D-85748
  Garching bei M\"unchen, Germany}

\maketitle 

\begin{abstract}
We present mid-IR observations of the prototypical Seyfert 2 galaxy
NGC\,1068 obtained with ISOCAM between 5 and 16\,$\mu$m. The proximity
of this galaxy coupled with the spectro-imaging capabilities of the
instrument allow us to separate the mid-IR emission of the active
galactic nucleus (AGN) from the distinctly different emission of the
circumnuclear star forming regions. The Unidentified Infrared Bands
(UIBs), which trace the starburst contribution very well, are not
detected in the spectrum of the AGN region where their carriers could
be destroyed. Moreover, the featureless continuum of the AGN exhibits
a strong hot dust component below 10\,$\mu$m not observed in the
starburst regions. Those two distinct mid-IR spectral properties, as
well as the presence of high excitation ionic lines such as
[NeVI](7.7\,$\mu$m) and [NeV](14.3\,$\mu$m) in the AGN spectrum,
provide us with very powerful complementary tools to disentangle AGNs
from starbursts.  The effects of high extinction on the mid-IR
identification of AGNs are also discussed.

\end{abstract}

\section{INTRODUCTION}
The understanding of energetic phenomena due to star formation and/or
an AGN observed in nearby galaxies is of crucial importance to explaining
the output luminosity produced by ultra-luminous infrared galaxies
(L$_{IR}$$>$10$^{12}$L$_{\odot}$, see \cite{Sanders} for a review).
Recent studies based on near-IR, mid-IR (\cite{Murphy}, \cite{Soifer},
\cite{Genzelb}), and X-ray (\cite{Risaliti}) observations indicate that
those galaxies are probably dominated by starbursts over their global
infrared luminosity, even though weak AGNs may still be present in most
of them. The possible detection of weak AGNs in all ultra-luminous
galaxies is still plausible and necessitates powerful diagnostics.

In this paper, we examine the mid-IR spectral properties observed in
the central region of the prototypical Seyfert 2 galaxy NGC 1068
(\cite{Lefloch}). Due to the proximity of this galaxy (14.4\,Mpc,
1$''$=72pc), we can disentangle the AGN from the star formation
regions found at 20$''$ from the nucleus. A comparative study of their
respective mid-IR emission will be presented in order to point out, in
a more general way, different methods for distinguishing an AGN within
a dominant starburst environment.

\section{OBSERVATIONS}
The galaxy NGC 1068 was observed with the ISOCAM camera
(\cite{Cesarsky}) on board the Infrared Space Observatory (ISO,
\cite{Kessler}). It is part of the active galaxy proposal CAMACTIV
(P.I. I.F. Mirabel), which contains 29 nearby galaxies hosting star
formation activity and AGN signatures (see \cite{Laurent}).  We have
used the circular variable filter (CVF) for mapping the central region
of NGC 1068 (96$''$$\times$96$''$) continuously from 5 to 16\,$\mu$m
with a spectral resolution of $\sim$\,40.  The data reduction was
performed with the CAM Interactive Analysis software (CIA\footnote{CIA
is a joint development by the ESA astrophysics division and the ISOCAM
consortium led by the ISOCAM PI, C. Cesarsky, Direction des Sciences
de la mati\`ere, C.E.A. France.})  and calibration followed the
general methods described in \cite*{Starck}. In addition, the jitter
effect due to the satellite motions (amplitude $\sim$ 0.5\,arcsec)
which severely affects the flux distribution of the bright point-like
nucleus was corrected using shift techniques. We also applied aperture
corrections to account for the overall extension of the PSF. The
absolute uncertainty of our photometric measurements is $\sim$\,20$\%$,
while the error in the relative flux uncertainty mainly due to
limitations on transient effect correction is $\sim$10$\%$.

\section{AGN AND STARBURST MID-IR EMISSION}

In Figure \ref{laurento1}, we present mid-IR spectra and images of
NGC\,1068 obtained with ISOCAM. The emission between 5 and 16\,$\mu$m
is clearly dominated by the AGN. Nevertheless, faint extended starburst
emission is observed along the northeast-southwest axis (see upper
and lower right panels). One may notice that the AGN and the faint
surrounding starburst have distinctly different mid-IR properties:

\begin{figure*}[!t]
\vspace{-15mm}
  \begin{center}
  \resizebox{16cm}{!}{\includegraphics{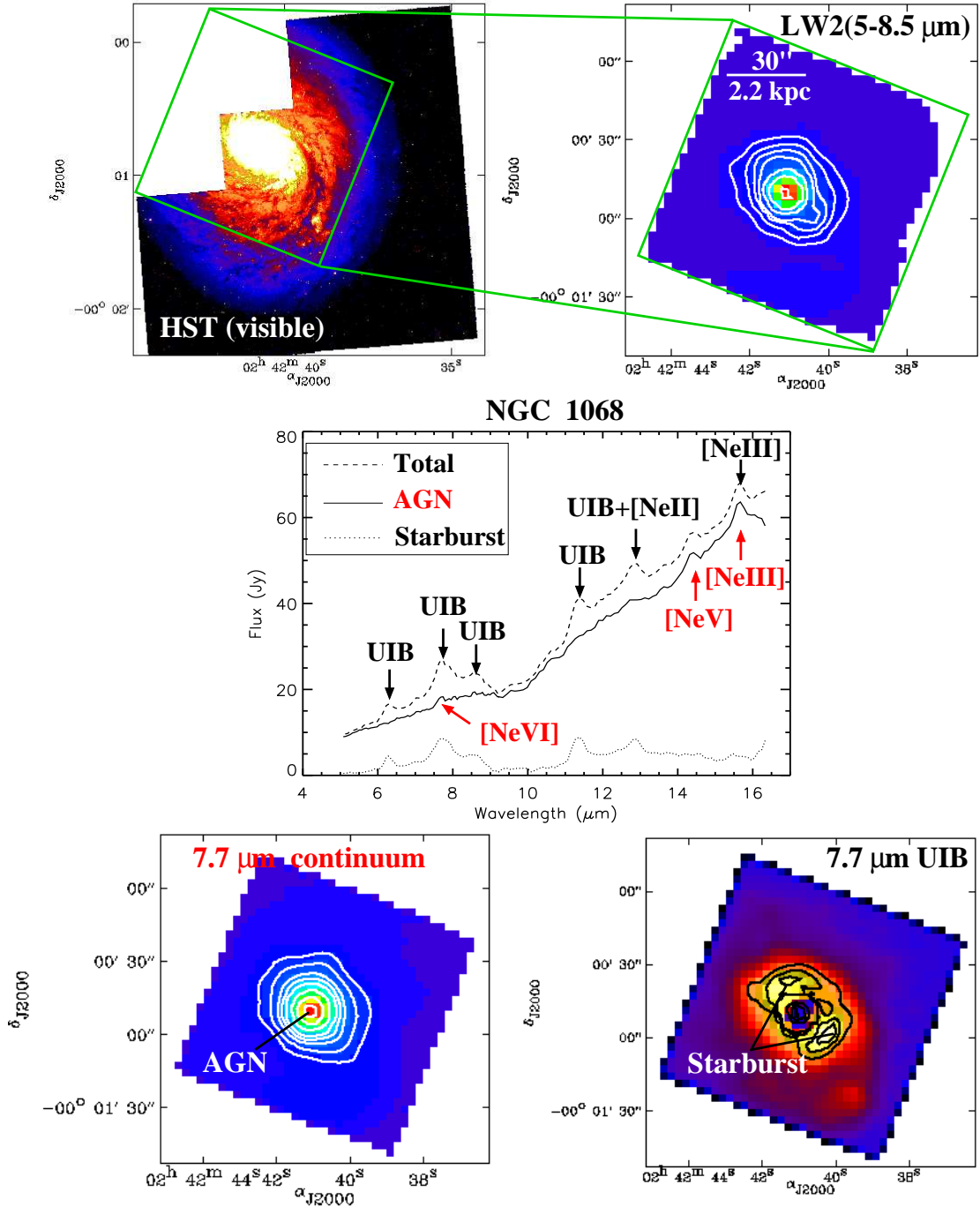}} 
\end{center}
\vspace{-2mm}
\caption{Mid-IR spectro-imaging observations of NGC 1068 with ISOCAM. 
Upper left panel\,: HST optical image (660\,nm) with the ISOCAM field
of view superimposed. Upper right panel\,: ISOCAM image and contours
in the 5-8.5\,$\mu$m band which traces the intense UIB emission at
7.7\,$\mu$m as well as the AGN continuum. Middle panel\,: Dashed
line\,: Total integrated spectrum of the galaxy ($\sim$40\,$''$,
29\,kpc in diameter). Thick line\,: Spectrum of the nuclear region
($\sim$9\,$''$, 0.7\,kpc in diameter) composed of a strong featureless
continuum and high excitation ionic lines ([NeV], [NeVI]). Dotted
line\,: Spectrum of the extended star forming regions
(40\,$''$$<$Diameter$<$9\,$''$) characterized by strong UIBs compared
to the continuum. Lower left panel\,: Image of the hot continuum
emission orginating essentially from the AGN, and defined as being under
the straight line between 7.3 and 8.2\,$\mu$m. Lower right panel\,:
Image of UIB emission above the straight line between 7.3 and
8.2\,$\mu$m, tracing the starburst regions. The contours are from the
deconvolved 7.7\,$\mu$m map (LW6(7-8.5\,$\mu$m) filter), which allows
us to separate the strong AGN contamination from the starburst
emission. Note that the two methods (spatial deconvolution and
7.7\,$\mu$m UIB map) used for spatially separating the faint starburst
emission are in good agreement.}
\label{laurento1}
\vspace{-3mm}
\end{figure*}

\begin{table*}[!t]
\begin{center}
\caption{Spectral properties of AGNs and starbursts in the mid-IR between 5 and 16\,$\mu$m.}
\renewcommand{\arraystretch}{1.4}
\setlength\tabcolsep{7.5pt}
\begin{tabular}{|l| l c c c c c| l c c c c c|}
\hline
 	&  \multicolumn{6}{c|}{STARBURST}
	&  \multicolumn{6}{c|}{AGN} \\
\hline
DUST	& \multicolumn{6}{l}{Very small grain continuum at 10-16$\mu$m originating}
	& \multicolumn{6}{|l|}{Hot dust continuum at 5-16$\mu$m from the dusty} \\
 	& \multicolumn{6}{l}{mainly from HII regions.}
	& \multicolumn{6}{|l|}{torus heated by the AGN radiation field.}\\
\hline 
UIBs	& \multicolumn{6}{l}{UIBS at 6.2, 7.7, 8.6, 11.3 and 12.7$\mu$m produced}
  	& \multicolumn{6}{|l|}{Absence of UIBs probably due to the destruction of}\\
	& \multicolumn{6}{l}{mainly in photo-dissociation regions surrounding} 
	& \multicolumn{6}{|l|}{their carriers by the intense UV/X-ray radiation field}\\
	& \multicolumn{6}{l}{the HII regions.}
	& \multicolumn{6}{|l|}{of the AGN.}\\
\hline
GAS	&	 & [ArII] & [NeII] & [ArIII] & [SIV] & [NeIII]
	&	 & [ArII] & ...    & [NeIII] & [NeV] & [NeVI]\\
	& Ep(eV) & 16 	  & 22     & 28      & 35    & 41
	& Ep(eV) & 16 	  & ...    & 41      & 97    & 126\\
        & $\lambda$ ($\mu$m) 
                 & 6.99   & 12.81  & 8.99    & 10.51 & 15.55
        & $\lambda$ ($\mu$m) 
                 & 6.99   & ...    & 15.55   & 14.32 & 7.65\\
\hline
\end{tabular}\\
\label{table}
\end{center}
\end{table*}

$\bullet$ The most striking difference is the absence of the family of
UIBs at 6.2, 7.7, 8.6, 11.3, and 12.7\,$\mu$m in the AGN spectrum.  The
UIBs originate mainly from photo-dissociation regions surrounding HII
regions, and as a consequence, they are naturally associated with the
starburst activity. This depletion of the UIB carriers may indicate
their destruction by the hard radiation field from the AGN, as  is
observed in well-resolved pure Galactic HII regions as well as in some
low-metallicity starburst galaxies. The relative decrease of UIB
strength can be used as a good indicator for distinguishing the AGN
from the starburst regions, which are composed of PDRs and HII regions
(\cite{Genzela}, \cite{Rigopoulou}).\\

$\bullet$ The AGN spectrum presents a noticeable continuum at short
wavelengths (5-10\,$\mu$m), commonly attributed to hot dust,
associated with the torus of molecular gas proposed in the unified
model (\cite{Pier}, \cite{Granato}, \cite{Genzelb}).  Some intense HII
regions present a similar continuum, but only in regions of a few
parsecs in size located close to hot stars with up to 10$^5$ the
solar radiation field (see \cite{Contursi}). Nevertheless, this
continuum has never been reported in starburst regions on large
spatial scales ($>$100\,pc), and thus its presence constitutes very
strong evidence for an AGN.\\

$\bullet$ The large variety of ionic emission lines, with ionization potential 
ranging from 22\,eV for [ArII] up to 126\,eV for [NeVI] can be
used to estimate the hardness of the radiation field (see Table
\ref{table}).  Highly ionized species tracing the hard radiation field
of the AGN (e.g. [NeV]14.3\,$\mu$m and [NeVI]7.6\,$\mu$m) are detected
in NGC\,1068. However, due to the low spectral resolution of ISOCAM
spectra ($\lambda / \Delta \lambda \sim$\,40), their detection in
other nearby AGNs is still difficult and requires higher resolution
(\cite{Sturm}).  The [NeV] emission line can also be induced by shocks
such as those created by supernova remnants (SNRs). The very weak
mid-IR continuum associated with SNRs, though, rules out the
possibility that there is a strong contribution of SNRs to the AGN
environment (\cite{Oliva}), and the detection of these lines still
represents the most direct way of identifying AGNs.

In Table \ref{table}, we summarize the overall properties of AGN
and starburst mid-IR emission between 5 and 16\,$\mu$m.  In the
following section, we discuss how dust extinction in starbursts and
AGNs may lead us to underestimate the AGN component.

\begin{figure*}[!t]
  \begin{center}
  \resizebox{\hsize}{!}{\includegraphics{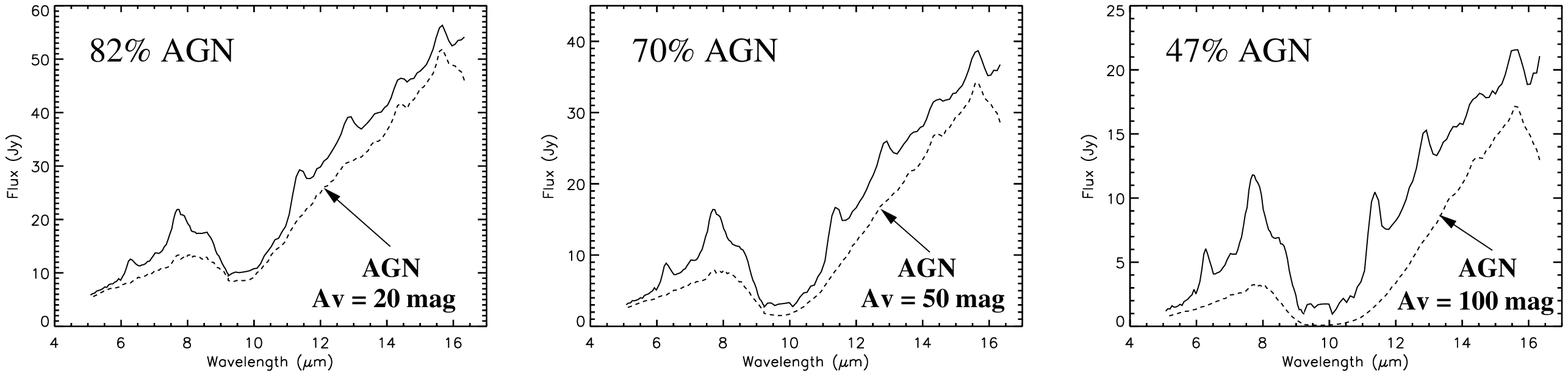}} \end{center}
  \caption{Extinction effects on the AGN continuum of NGC\,1068. From
  left to right, the AGN continuum is reddened by 20, 50 and 100 mag
  respectively, using a screen model with the extinction law of Lutz
  et al. (1996). Note how the estimated contribution of the AGN to the
  total mid-IR decreases and the galaxy evolves from an AGN-dominated
  (87$\%$) to a starburst-dominated system (47$\%$).}
\label{laurento2}
\end{figure*}

\section{THE EFFECTS OF EXTINCTION}

According to the unified picture of AGNs proposed by \cite*{Antonucci},
the central engine (a massive black hole accreting material) 
is surrounded by a dusty molecular torus.  The additional absorption
caused by this torus would normally decrease the intrinsic luminosity
of the AGN compared to the adjacent starburst regions. This leads to
the Seyfert 2/Seyfert 1 optical classification, depending on whether
the torus is observed edge-on or face-on.  Studying a sample of 28
Seyfert 1 and 29 Seyfert 2 galaxies with ISOPHOT-S, Schulz et
al. (1998) have already shown that the high absorption in Seyfert 2
galaxies blocks a large fraction (90$\%$ on average) of the mid-IR
continuum from the AGN inner torus.  In Figure \ref{laurento2}, we
show the consequences for the mid-IR spectrum if additional
extinction is applied to the AGN continuum of NGC\,1068. As the
extinction increases, the continuum at 5-10\,$\mu$m, used as a good
AGN indicator, becomes less detectable than the UIB emission, causing
us to classify this galaxy as dominated by a starburst above
A$_v$=100.

\section{SIRTF/IRS PERSPECTIVES}

\begin{figure}[!ht]
\vspace{-5mm}
\hspace{-3mm}
  \resizebox{\hsize}{!}{\includegraphics{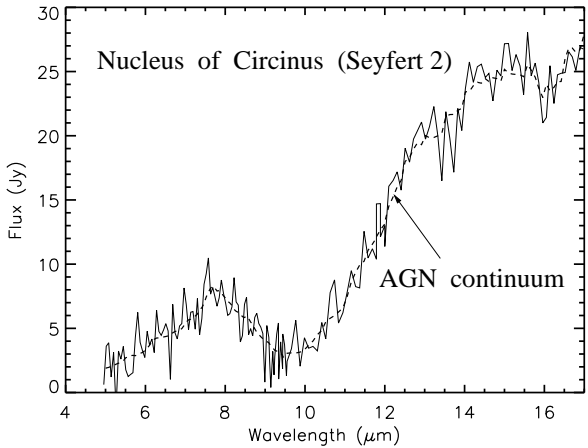}} 
\caption{ISOCAM spectrum of the AGN in Circinus (Moorwood (1999),
dashed line, 4.5$''$$\times$4.5$''$).  Note how the strong silicate
absorption makes difficult the distinction between an obscured AGN
continuum and a strong UIB feature at 7.7\,$\mu$m, in particular for an
observation with a low signal to noise (solid line, $\sigma_{noise}$=1 Jy)}
\label{laurento3}
\vspace{-10mm}
\end{figure}

Thanks to ISO, our knowledge of the mid-IR properties of starbursts
and AGNs has been substantially improved through observations of
nearby prototypical objects. In particular, different diagnostics were
developed to obtain a quantitative classification of starburst and
AGN-dominated galaxies using their mid-IR emission. A faint UIB to
continuum ratio (\cite{Genzela}) and the detection of a strong
continuum at 5\,$\mu$m (\cite{Lutzb}, \cite{Laurent}) are only
observed in galaxies containing an AGN. Nevertheless, the accuracy of
the estimate depends strongly on the signal to noise of the spectra
(see for example Fig. \ref{laurento3}). Another diagnostic for
distinguishing AGNs from starbursts is the ratios of high to low
excitation ionic lines, such as [NeV]/[NeII] or [OIV]/[NeII], which
probe the hardness of the radiation field (\cite{Genzela}).  However,
the detection of the [NeV] and [OIV] emission lines with ISO-SWS,
which directly traces the presence of the AGN, is still limited to
very nearby luminous objects. SIRTF, the next infrared space
telescope, and in particular its mid-IR spectrograph IRS
(\cite{Houck}), which is expected to be nearly 400 times more
sensitive than ISO-SWS, will be able to examine the possible presence
of weak AGNs in luminous galaxies. The use of emission line ratios
such as [SIII]18.7\,$\mu$m/[SIII]33.5\,$\mu$m (\cite{Genzela}) or the
7.7UIB/6.2UIB ratio (\cite{Rigopoulou}) will also provide us other
independent estimates of absorption in a larger sample of active
galaxies.  Furthermore, the full coverage from 5.3 to 40\,$\mu$m will
permit us to better estimate the effects of the absorption using the
9.7 and 18\,$\mu$m silicate features as well as reveal the presence of
a faint hot dust contribution associated with intrinsically weak or
obscured AGNs.

\begin{acknowledgements}
We wish to thank A.J. Baker for very useful suggestions.
\end{acknowledgements}

\end{document}